\documentclass[sigconf]{acmart}

\usepackage{graphicx}
\usepackage{epsfig}
\usepackage{subfig}
\usepackage{booktabs}
\usepackage{amsmath}
\usepackage{multirow}
\usepackage{bm}
\usepackage{hyperref}
\usepackage{balance}

\long\def\comment#1{}

\newcommand{\etal}{\textit{et al}.}
\newcommand{\ie}{\textit{i}.\textit{e}.}
\newcommand{\eg}{\textit{e}.\textit{g}.}

\newcommand{\VX}{\mathbf{X}}
\newcommand{\VY}{\mathbf{Y}}

\newcommand{\Vx}{\mathbf{x}}
\newcommand{\Vy}{\mathbf{y}}

\DeclareMathOperator*{\argmax}{argmax}

\def\BibTeX{{\rm B\kern-.05em{\sc i\kern-.025em b}\kern-.08emT\kern-.1667em\lower.7ex\hbox{E}\kern-.125emX}}
    
\copyrightyear{2021}
\acmYear{2021}
\setcopyright{acmcopyright}\acmConference[MM '21]{Proceedings of the 29th ACM International Conference on Multimedia}{October 20--24, 2021}{Virtual Event, China}
\acmBooktitle{Proceedings of the 29th ACM International Conference on Multimedia (MM '21), October 20--24, 2021, Virtual Event, China}
\acmPrice{15.00}
\acmDOI{10.1145/3474085.3475419}
\acmISBN{978-1-4503-8651-7/21/10}

\begin{document}

\fancyhead{}

%
\title[Locally Adaptive Structure and Texture Similarity]{Locally Adaptive Structure and Texture Similarity \\for Image Quality Assessment}

%

\author{Keyan Ding$^1$, Yi Liu$^2$, Xueyi Zou$^2$, Shiqi Wang$^1$, Kede Ma$^1$}
\affiliation{
  \institution{$^1$City University of Hong Kong, Hong Kong}
  \city{$^2$Noah's Ark Lab, Huawei Technologies, Shenzhen, China} \\
  \country{keyan.ding@my.cityu.edu.hk, \{liuyi113, zouxueyi\}@huawei.com, \{shiqwang, kede.ma\}@cityu.edu.hk}
}


%

%
\begin{abstract}
The latest advances in full-reference image quality assessment (IQA) involve unifying structure and texture similarity based on deep representations. The resulting Deep Image Structure and Texture Similarity (DISTS) metric, however, makes rather \textit{global} quality measurements, ignoring the fact that natural photographic images are \textit{locally} structured and textured across space and scale. In this paper, we describe a locally adaptive structure and texture similarity index for full-reference IQA, which we term A-DISTS. Specifically, we rely on a single statistical feature, namely the dispersion index, to localize texture regions at different scales. The estimated probability (of one patch being texture) is in turn used to adaptively pool local structure and texture measurements. The resulting A-DISTS is adapted to local image content, and is free of expensive human perceptual scores for supervised training. We demonstrate the advantages of A-DISTS in terms of \textit{correlation} with human data on ten IQA databases and \textit{optimization} of single image super-resolution methods. 
\end{abstract}


\maketitle

\section{Introduction}
Full-reference image quality assessment (IQA) aims to predict the perceived quality of a ``distorted'' image with reference to its original undistorted counterpart. It plays an indispensable role in the assessment and optimization of various image processing and computational photography algorithms. Humans appear to assess image quality quite easily and consistently, but the underlying mechanism is unclear, making bio-inspired IQA model development a challenging task. 

For more than half a century, the field of full-reference IQA has been dominated by parsimonious knowledge-driven models with few hyperparameters. Representative examples include the mean squared error (MSE), the structural similarity (SSIM) index~\cite{wang2004image}, the visual information fidelity (VIF) metric~\cite{sheikh2006image}, the most apparent distortion (MAD) measure \cite{larson:011006}, and the normalized Laplacian pyramid distance (NLPD)~\cite{laparra2016perceptual}. Knowledge-driven IQA methods require statistical modeling of the natural image manifold and/or the human visual system (HVS) \cite{duanmu2021iqa}, which is highly nontrivial. Only crude computational approximations characterized by simplistic and restricted visual stimuli \cite{watson2000visual} have been developed. 

Recently, there has been a trend shying away from knowledge-driven IQA models and toward data-driven ones, as evidenced by recent IQA methods~\cite{bosse2018deep,zhang2018unreasonable,prashnani2018pieapp,ding2020dists} based on deep neural networks (DNNs). Albeit with high correlation numbers on many image quality databases, these models have a list of theoretical and practical issues during deployment  \cite{wang2016objective}. Arguably the most significant issue is with regard to gradient-based optimization. In \cite{ding2020optim}, Ding \etal~systematically evaluated more than $15$ full-reference IQA models in the context of perceptual optimization, and found that a majority of methods fail in a na\"{i}ve task of reference image recovery. This is not surprising because these methods rely on \textit{surjective} mapping functions to transform the original and distorted images to a reduced ``perceptual'' space for quality computation~\cite{ding2020optim}. 
Two DNN-based models that rely on (nearly) \textit{injective} mappings are the exceptions: the learned perceptual image patch similarity (LPIPS)~\cite{zhang2018unreasonable} and the deep image structure and texture similarity (DISTS) \cite{ding2020dists} metrics. According to the subjective user study in \cite{ding2020optim}, LPIPS and DISTS are top-$2$ performers in optimization of three low-level vision tasks - blind image deblurring, single image super-resolution, and lossy image compression (see Fig. 8 in \cite{ding2020optim}). 

\begin{figure}
  \centering
    \subfloat[]{\includegraphics[height=0.375\linewidth]{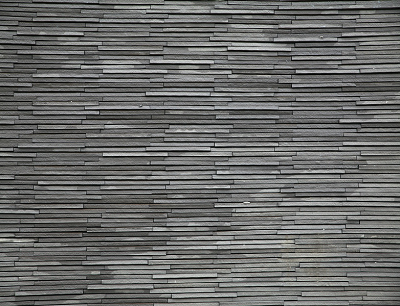}}\hskip.3em
    \subfloat[]{\includegraphics[height=0.375\linewidth]{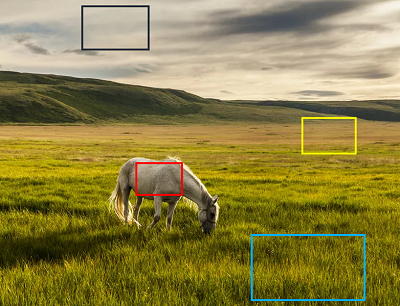}}
  \caption{Visual comparison between (a) a global texture image and (b) a natural photographic image. We can observe in (b) that texture elements of different visual appearances are distributed at different locations and scales, as indicated by the bounding boxes.}
  \label{fig:natural_texture}
\end{figure}

LPIPS \cite{zhang2018unreasonable} makes quality measurements by point-by-point comparisons between deep features from the pre-trained VGG network \cite{Simonyan14c}. As a result, it cannot properly handle ``visual texture,'' which is comprised of repeated patterns, subject to some randomization in their location, size, color, and orientation \cite{portilla2000parametric}. DISTS~\cite{ding2020dists} provides a better account for texture similarity by identifying a compact set of statistical constraints as global spatial averages of VGG feature maps. When restricted to \textit{global} texture images (see Fig. \ref{fig:natural_texture} (a)), the underlying texture model in DISTS performs well in the analysis-by-synthesis test \cite{portilla2000parametric} originally advocated by Julesz \cite{julesz1962visual}. However, it is well-known that natural photographic images are composed of ``things'' (\ie, objects) and ``stuff'' (\ie, textured surfaces) \cite{adelson2001seeing}, \textit{localized} in space and scale (see Fig. \ref{fig:natural_texture} (b)). Therefore, it is desirable to develop structure and texture similarity methods adapted to local image content.

In this paper, we propose a \textit{locally adaptive} DISTS metric, which we name A-DISTS for full-reference IQA. The central idea is to compute a spatially varying  map at a certain scale (\ie, a convolution stage in VGG \cite{Simonyan14c}), where each entry indicates the probability of the patch within the receptive field being texture. We identify a single statistical feature, the dispersion index \cite{Cox1966} computed as the variance-to-mean ratio of convolution responses, which provides an excellent separation between structure and texture patches. The probability map is in turn used to adaptively pool local structure and texture similarity measurements across spatial locations and convolution channels. The resulting A-DISTS is adapted to local image content, and is free of expensive mean opinion scores (MOSs) for supervised training. Our extensive experiments based on ten human-rated IQA databases \cite{LIVE,larson:011006,Ponomarenko201557,lin2019kadid,zhang2018unreasonable,ding2020dists,lai2016comparative,ma2017learning,min2019quality,tian2018benchmark} show that A-DISTS leads to consistent performance improvements in terms of correlation with MOSs, especially on datasets with distortions arising from real-world image restoration applications. Moreover, A-DISTS demonstrates competitive performance in perceptual optimization of single image super-resolution methods.

\section{Related Work}
This section reviews four full-reference IQA models that are relevant to the proposed A-DISTS:  MSE,  SSIM \cite{wang2004image},  LPIPS \cite{zhang2018unreasonable}, and DISTS \cite{ding2020dists}. The former two have made a profound impact on a wide range of multimedia signal processing algorithms, while the later two rely on the same deep feature representation as A-DISTS. All four models have been proven effective in the context of perceptual optimization \cite{ding2020optim}.

We use bold capital letters such as $\VX$ and $\VY$ to represent input images, bold lower-case letters such as $\Vx_k$ and $\Vy_k$ to represent the $k$th input image patches, and lower-case letters such as $x_k$ and $y_k$ to represent the $k$th pixel values. Similarly, we use bold capital letters with tildes such as $\tilde{\VX}^{(i)}_j$ and $\tilde{\VY}^{(i)}_j$ to represent the convolution response (\ie, the feature map) from the $j$th channel of the $i$th stage of a DNN (corresponding to $\VX$ and $\VY$), bold lower-case letters with tildes such as $\tilde{\Vx}^{(i)}_{j,k}$ and $\tilde{\Vy}^{(i)}_{j,k}$ to represent the $k$th feature patches (in $\tilde{\VX}^{(i)}_j$ and $\tilde{\VY}^{(i)}_j$). \comment{and lower-cases letters with tildes such as $\tilde{x}^{(i)}_{j,k}$ and $\tilde{y}^{(i)}_{j,k}$ to represent the $k$th feature coefficient.} We denote by $\mathcal{X}$ and $\tilde{\mathcal{X}}$ the image and feature spaces, respectively. We use  $f:\mathcal{X}\mapsto\tilde{\mathcal{X}}$ to represent the feature transform which maps an input image to a perceptually plausible representation. It can be the identity mapping (\ie, $\VX = \tilde{\VX}$) or parameterized by DNNs.

\subsection{MSE}
MSE is defined as the average of the squares of the errors between the original undistorted image $\VX$ and the test ``distorted'' image $\VY$:
\begin{equation}
\mathrm{MSE}(\VX, \VY)=\frac{1}{K}\sum_{k=1}^K \left(x_{k}-y_{k}\right)^{2},
\end{equation}
where $K$ is the number of pixels in the image. It is simple, physically plausible (\eg, energy preserving according to the Parseval's theorem), and mathematically beautiful. For example, when MSE is combined with Gaussian source and noise models, the optimal solution in the signal estimation framework is  analytical and linear \cite{wang2009mean}. 
One major drawback of MSE and its derivatives is their poor correlation with human perception of image quality.

\begin{figure*}[t]
  \centering
    \subfloat[]{\includegraphics[height=0.19\linewidth]{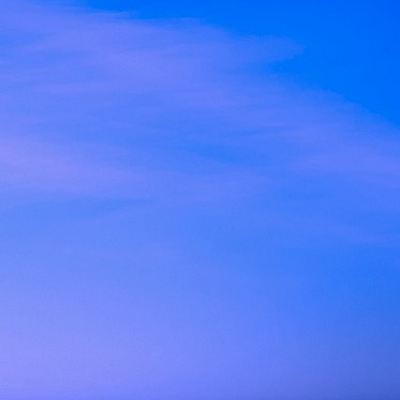}}\hskip.3em
    \subfloat[]{\includegraphics[height=0.19\linewidth]{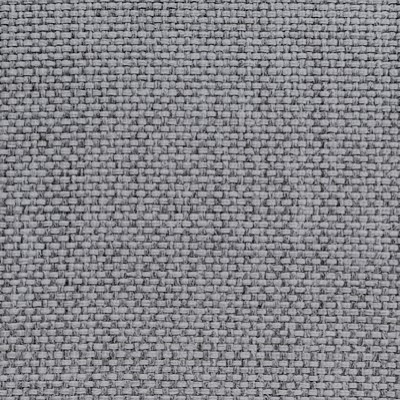}}\hskip.3em
    \subfloat[]{\includegraphics[height=0.19\linewidth]{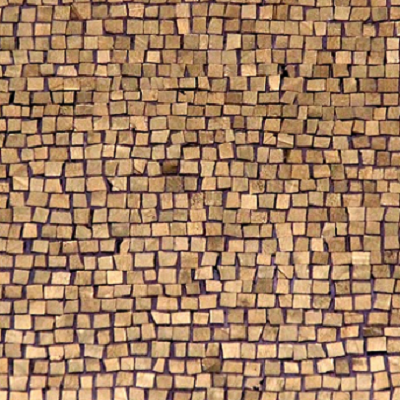}}\hskip.3em
    \subfloat[]{\includegraphics[height=0.19\linewidth]{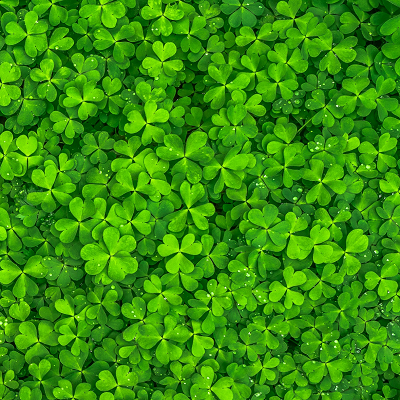}}\hskip.3em
    \subfloat[]{\includegraphics[height=0.19\linewidth]{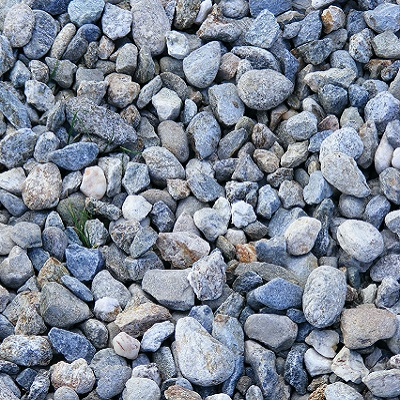}}
  \caption{Human perception of visual texture is scale-dependent. (a) - (e): Texture images with increasing scales. Images (a) and (b) allow a relatively small receptive field to capture the intrinsic repetitiveness. In contrast, images (d) and (e) that are composed of small-scale textured surfaces and structural contours require a large receptive field to sufficiently cover the repeated patterns.
  }
  \label{fig:tex_5scale}
\end{figure*}

\subsection{SSIM}
SSIM \cite{wang2004image} is a top-down approach, motivated by the observation that natural photographic images are highly structured. Therefore, a measure of the retention of local image structure should provide a reasonable approximation to perceived image quality. Regardless of various instantiations, the basic form of SSIM measures the similarities of local patch intensities, contrasts, and structures. These are computed using simple patch statistics, and can be combined and simplified to
\begin{equation}\label{eq:lssim}
\operatorname{SSIM}(\Vx, \Vy)=l(\Vx,\Vy)\cdot s(\Vx,\Vy)=\left(\frac{2 \mu_{\Vx} \mu_{\Vy}+c_{1}}{\mu_{\Vx}^{2}+\mu_{\Vy}^{2}+c_{1}}\right)\cdot\left(\frac{2 \sigma_{\Vx \Vy}+c_{2}}{\sigma_{\Vx}^{2}+\sigma_{\Vy}^{2}+c_{2}}\right),
\end{equation}
where $\mu_{\Vx}$ and $\mu_{\Vy}$ are (respectively) the mean intensities of $\Vx$ and $\Vy$, $\sigma_{\Vx}^{2}$ and $\sigma_{\Vy}^{2}$ are (respectively) the variances of $\Vx$ and $\Vy$, and $\sigma_{\Vx \Vy}$ is the covariance between $\Vx$ and $\Vy$. $c_1$ and $c_2$ are two small positive constants, preventing potential division by zero. The local SSIM values are computed using a sliding window approach, and are averaged spatially to obtain an overall quality score:
\begin{align}
    \operatorname{MSSIM}(\VX, \VY) =  \frac{1}{K}\sum_{k=1}^K\operatorname{SSIM}(\Vx_k,\Vy_k),
\end{align}
where we slightly abuse $K$ to denote the number of (overlapping) patches in the image. It is widely acknowledged that SSIM is better at explaining human perceptual data than MSE. Nevertheless, Ding \etal~\cite{ding2020optim} found surprisingly that the perceptual gains of the multi-scale version of SSIM \cite{wang2003multiscale} over MAE are statistically indistinguishable when optimizing four low-level vision tasks.

\subsection{LPIPS}
LPIPS \cite{zhang2018unreasonable} leverages the ``unreasonable'' effectiveness of deep features to account for many aspects of human perception, and computes a weighted MSE between normalized feature maps of two images:
\begin{equation}
\operatorname{LPIPS}\left(\VX, \VY\right) =  \sum_{i=1}^{M}\sum_{j=1}^{N_i}  w_{ij}\operatorname{MSE}\left(\tilde{\VX}_j^{(i)}, \tilde{\VY}_j^{(i)}\right),
\label{eq:lpips}
\end{equation}
where $w_{ij}\ge 0$ is learnable, indicating the perceptual importance of each channel. $M$ denotes the number of convolution stages, and $N_i$ is the number of feature maps in the $i$th stage. LPIPS has multiple configurations, and we adopt the one based on the VGG network~\cite{Simonyan14c} with weights learned from the BAPPS dataset~\cite{zhang2018unreasonable} throughout the paper.
It is noteworthy that VGG-based LPIPS can be seen as a generalization of the ``perceptual loss'' \cite{johnson2016perceptual}, which is widely used in image restoration tasks.

\subsection{DISTS}
DISTS \cite{ding2020dists} is based on a variant of the VGG network, and makes global SSIM-like structure and texture similarity measurements: 
\begin{align}
\operatorname{DISTS}(\VX, \VY)=1-\sum_{i=0}^{M}\sum_{j=1}^{N_i}\left(\alpha_{ij}l\left(\tilde{\VX}^{(i)}_j,\tilde{\VY}^{(i)}_j\right)+ \beta_{ij}s\left(\tilde{\VX}^{(i)}_j,\tilde{\VY}^{(i)}_j\right)\right),
\label{eq:vgg-metric}
\end{align}
where $\{\alpha_{ij}, \beta_{ij}\}$ are positive learnable weights, optimized to match human perception of
image quality and invariance to resampled texture patches \cite{ding2020dists}. Some key modifications of DISTS relative to SSIM and LPIPS are worth mentioning. First, $\ell_2$-pooling is adopted to replace the max pooling in the original VGG, which is conducive to de-alias and linearize the intermediate representations  \cite{henaff2015geodesics}. Second, the input image is incorporated as an additional feature map (\ie,  $\tilde{\VX}^{(0)} = \VX$) to guarantee the injectivity of the feature transform $f$. Third, unlike Eq. \eqref{eq:lssim}, DISTS applies the ``texture'' similarity function $l(\cdot)$ and the structure similarity function $s(\cdot)$ globally to compare feature maps. It has been empirically proven sensitive to structural distortions and robust to texture substitutions.

\section{A-DISTS}
In this section, we present in detail the locally adaptive DISTS metric, namely A-DISTS. We first describe the use of the dispersion index to separate structure and texture at different locations and scales. We then compute the final quality score by adaptively weighting local structure and texture measurements.

\vspace{2mm}\noindent\textbf{Structure and Texture Separation.} We want to identify robust statistics based on deep representations that are effective in separating structure and texture regions. However, the VGG network~\cite{Simonyan14c} used in DISTS suffers from the scale ambiguity. That is, we may re-scale a convolution filter by dividing the 3D tensor (and the associated bias term) by an arbitrary non-zero scalar. This can be compensated by re-scaling the next convolution filter connected to it by the same amount without changing the final softmax output. The scale ambiguity arises primarily from the adoption of half-wave rectification (\ie, ReLU) as the nonlinearity. As a consequence, the statistics computed from different convolution responses may be of arbitrary scale. To resolve this, we re-normalize the convolution filters (with size of $\mathrm{height}\times \mathrm{width} \times \mathrm{in\_channel}$) in VGG such that the $\ell_2$ norm of each filter is equal to one. With such re-normalization, all convolution filters have responses with similar ranges, making the computed statistics more comparable. Gatys \etal~\cite{gatys2015texture} noticed the same issue, and used a different form of re-scaling such that the average response of each filter over spatial locations and channels is equal to one.

\begin{figure*}[t]
  \centering
    \subfloat[]{\includegraphics[height=0.32\linewidth]{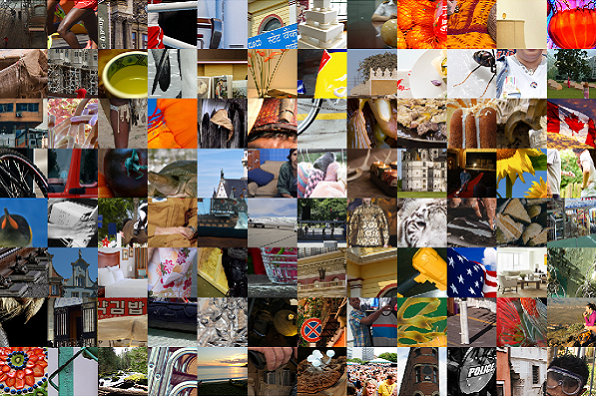}}\hskip.5em
    \subfloat[]{\includegraphics[height=0.32\linewidth]{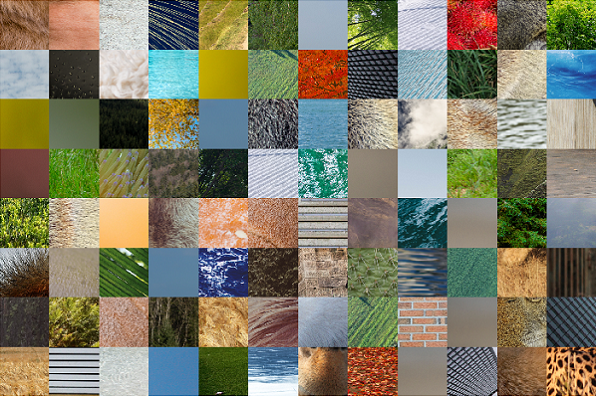}}
  \caption{Sample (a) structure and (b) texture patches of size 128$\times$128 in our image patch dataset manually cropped from the Waterloo Exploration Database \cite{ma2016waterloo} and the DIV2K dataset \cite{timofte2017ntire}.}
  \label{fig:sample_s_t}
\end{figure*}

\begin{figure*}[t]
  \centering
    \subfloat[$\gamma_{\Vx}^{(1)}$]{\includegraphics[height=0.17\linewidth]{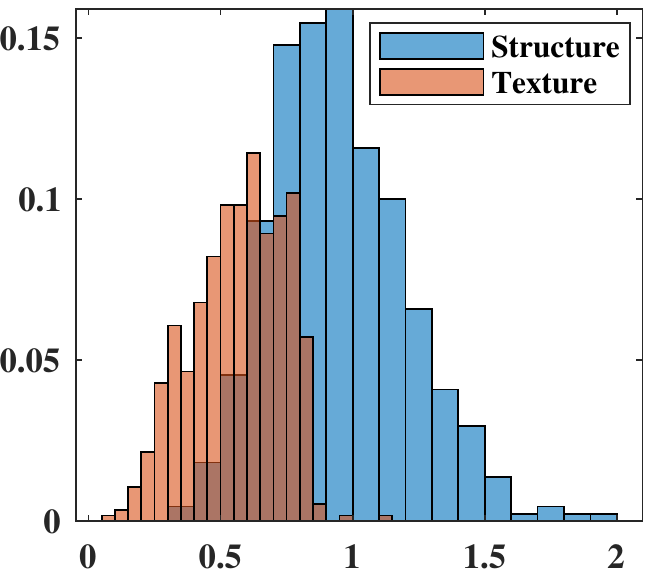}}\hskip.3em
    \subfloat[$\gamma_{\Vx}^{(2)}$]{\includegraphics[height=0.17\linewidth]{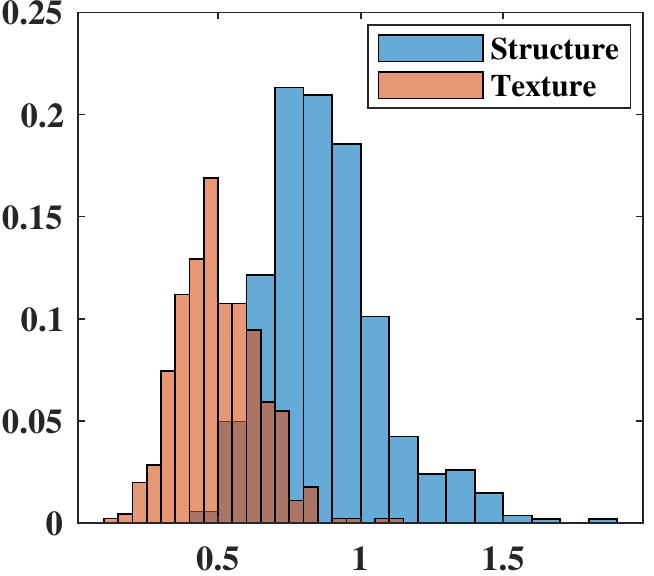}}\hskip.3em
    \subfloat[$\gamma_{\Vx}^{(3)}$]{\includegraphics[height=0.17\linewidth]{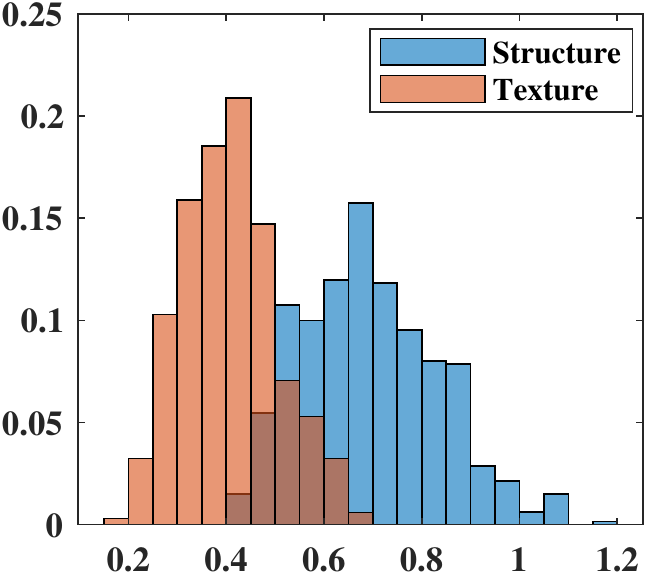}}\hskip.3em
    \subfloat[$\gamma_{\Vx}^{(4)}$]{\includegraphics[height=0.17\linewidth]{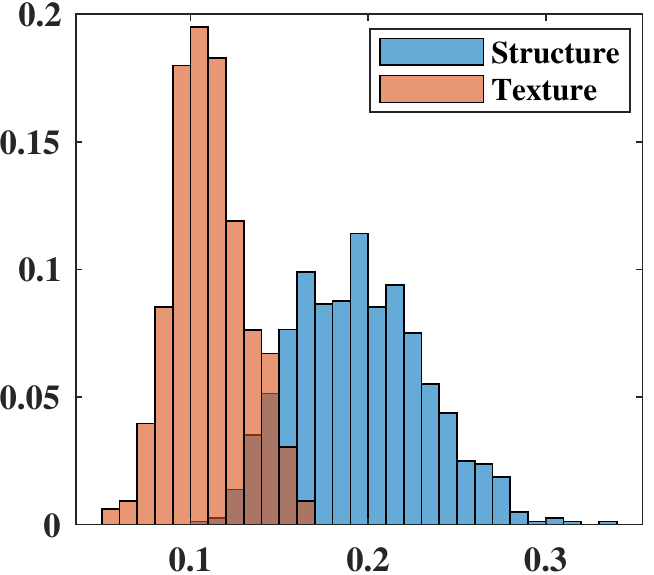}}\hskip.3em
    \subfloat[$\gamma_{\Vx}^{(5)}$]{\includegraphics[height=0.17\linewidth]{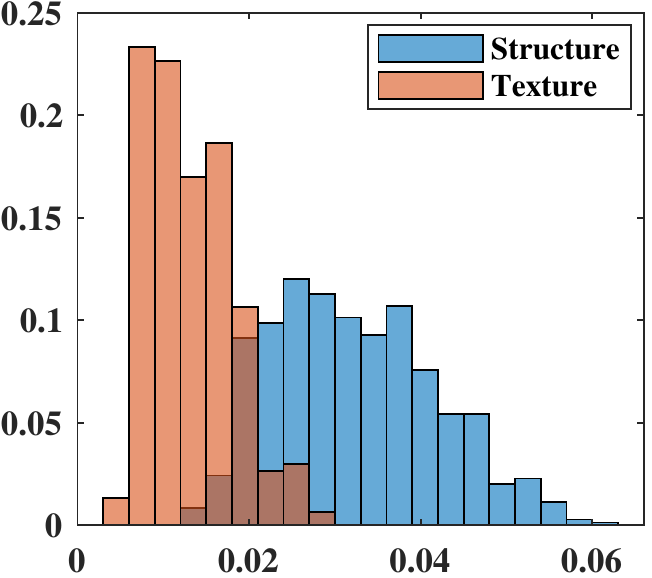}}
  \caption{The conditional histograms (normalized to probabilities) of the dispersion index $\gamma_{\Vx}^{(i)}$. One can observe that a clear separation between structure and texture at different stages is achieved.
  }
  \label{fig:distri_ratio}
\end{figure*}

We achieve the discrimination of structure and texture by exploiting two distinct characteristics. First, texture is spatially homogeneous, while structure is more precisely localized in space. Second, the perception of visual texture is scale-dependent. For small-scale visual texture (see Fig. \ref{fig:tex_5scale} (a) and (b)), a small receptive field (\eg, a $16\times16$ window) is able to capture its intrinsic repetitiveness, while for large-scale visual texture that is a combination of small-scale textured surfaces and structural contours  (see Fig. \ref{fig:tex_5scale} (d) and (e)), a large receptive field (\eg, a $128\times128$ window) may be needed to sufficiently cover the repeated patterns. 
Computationally, we use the dispersion index \cite{Cox1966} defined by the ratio of variance to mean as the structure/texture indicator. For each stage of VGG,  we apply a sliding window approach to compute local dispersion indexes, followed by averaging across channels:
\begin{align}
\gamma_{\Vx}^{(i)}=\frac{1}{N_i} \sum_{j=1}^{N_i} \frac{\left(\sigma_{\tilde{\Vx}_j}^{(i)}\right)^{2}}{\mu_{\tilde{\Vx}_j}^{(i)}+c},
\label{eq:ratio}
\end{align}
where $\mu_{\tilde{\Vx}_j}^{(i)}$ and $\sigma_{\tilde{\Vx}_j}^{(i)}$ represent (respectively) the mean and standard deviation of the local feature patch $\tilde{\Vx}^{(i)}_j$ in $\tilde{\VX}_j^{(i)}$. $c$ is a small positive stabilizing constant. The average operation is legitimate due to the re-normalization of the convolution filters in VGG. Intuitively, texture is often under-dispersed compared with structure, leading to a smaller $\gamma_{\Vx}^{(i)}$. As the receptive field of the VGG increases with the number of convolution and sub-sampling layers, we expect an early-stage $\gamma_{\Vx}^{(i)}$ is responsive to small-scale texture, while a late-stage $\gamma_{\Vx}^{(i)}$ is responsible for large-scale texture. To verify this, we gather an image patch dataset, which contains $2,500$ structure patches and $2,500$ texture patches of five different sizes (\ie, $16\times 16$, $32\times 32$, $64\times 64$, $128\times 128$, and $256\times 256$).  All patches are cropped from the Waterloo Exploration Database \cite{ma2016waterloo} and the DIV2K dataset \cite{timofte2017ntire}, and manually labeled. Fig. \ref{fig:sample_s_t} shows sample patches of size $128\times 128$, where we see great variability in structure arrangements and texture appearances. We draw the conditional histograms in Fig. \ref{fig:distri_ratio}, where we find a clear separation between structure and texture at different scales.

\begin{figure*}[t]
  \centering
    \subfloat[]{\includegraphics[height=0.26\linewidth]{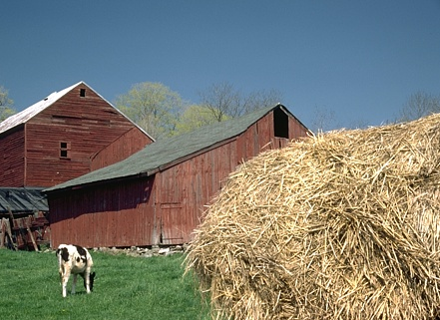}}\hskip.3em
    \subfloat[]{\includegraphics[height=0.26\linewidth]{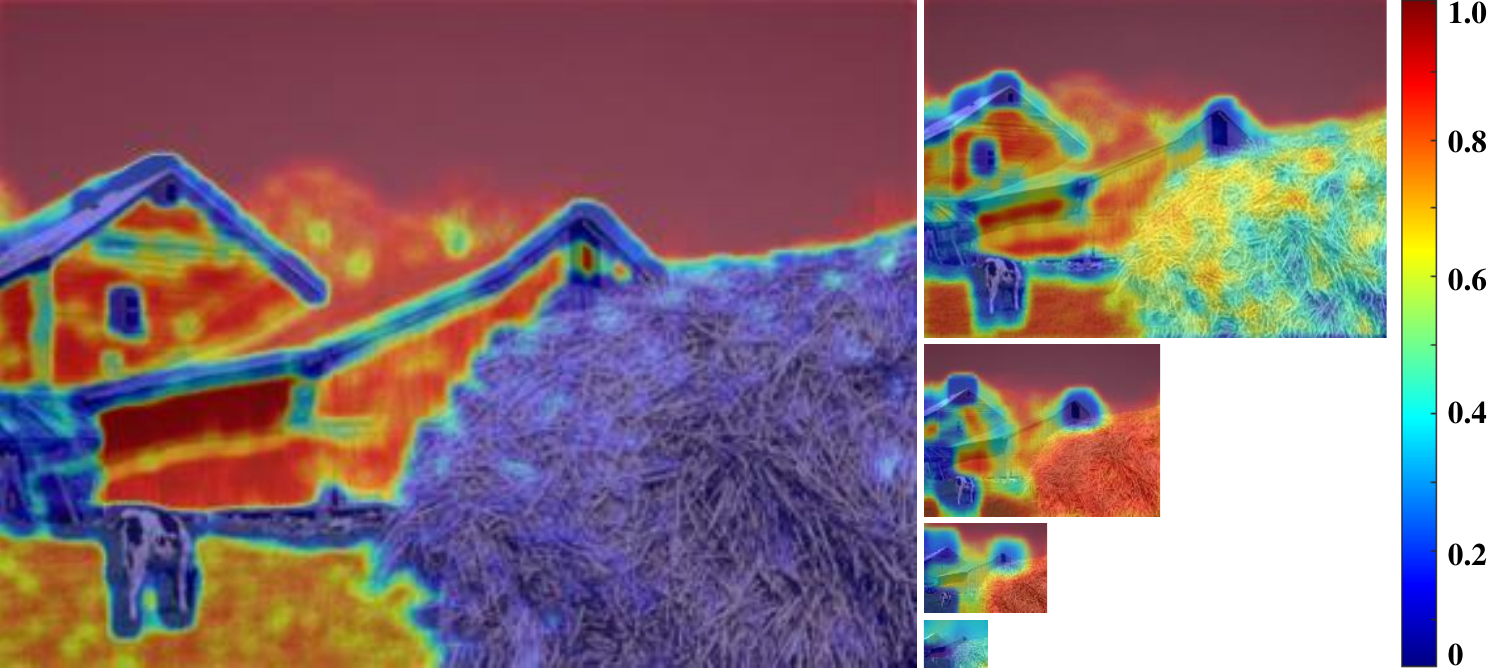}}
  \caption{Illustration of multi-scale texture probability maps. (a): Original image. (b): Texture probability maps with decreasing spatial sizes due to sub-sampling. The warmer the color is, the higher texture probability of the given patch is.}
  \label{fig:ps_map}
\end{figure*}

We then feed the dispersion index $\gamma_{\Vx}^{(i)}$ as a single statistical feature to logistic regression to compute the probability of the given patch being texture:
\begin{equation}
p_{\Vx}^{(i)}=p\left(``\Vx \mbox{ is texture}" \mid \gamma_{\Vx}^{(i)}\right)=\frac{1}{1+e^{-\left(w^{(i)} \gamma_{\Vx}^{(i)}+b^{(i)}\right)}},
\label{eq:pt}
\end{equation}
where $w^{(i)}, b^{(i)}$ are the weight and bias parameters to be fitted on our image patch dataset. 


Fig. \ref{fig:ps_map} shows the multi-scale texture probability maps of the ``Farm'' image, where a warmer color indicates a higher texture probability. Let us focus on the ``hay'' in the bottom right of the image. When we rely on $\gamma_{\Vx}^{(1)}$ that uses a small receptive field, the hay is classified as rather isolated structure, as reflected in the probability map at the finest scale. When we increase the receptive field (\eg, using $\gamma_{\Vx}^{(3)}$ or $\gamma_{\Vx}^{(4)}$), the hay is identified as texture, where the intrinsic repetitiveness is well captured. If we continue increasing the receptive field, the bottom right region containing the hay is classified towards structure again, which makes perfect sense because the receptive field is large enough to include surrounding structural contours (\eg, the boundaries of the hay and the farmhouse). Other small-scale texture such as the sky, the meadow, and the roof has also been successfully captured by $\gamma_{\Vx}^{(i)}$ and well reflected in the corresponding probability maps. Similarly, when the receptive field is large enough to include object boundaries, the included texture is part of the structure patch.

\vspace{2mm}\noindent\textbf{Perceptual Distance Metric.} With the multi-scale texture probability maps at hand, we are ready to design the spatial pooling strategy for combining local structure and texture similarity measurements.  As the proposed quality model is full-reference, we are able to compute two set of probability maps, $\{p_{\Vx}^{(i)}\}_{i=1}^5$ and $\{p_{\Vy}^{(i)}\}_{i=1}^5$ from the VGG representations of the co-located reference and distorted patches, $\Vx$ and $\Vy$, respectively. For the purpose of quality assessment, we take the minimum of the two texture probabilities:
\begin{align}
  \tilde{p}^{(i)} = \min\left(p^{(i)}_{\Vx} , p^{(i)}_{\Vy}\right), 
\end{align}
which is conductive to penalizing the introduced structural artifacts (\eg, JPEG blocking). For the purpose of perceptual optimization (as described in Section \ref{sec:sr}), we may directly use $p^{(i)}_{\Vx}$ as weighting to full respect the reference information.

Finally, we define the A-DISTS index as  
\begin{align}
\textrm{A-DISTS}(\VX, \VY)=1 - \frac{1}{N} \sum_{i=0}^{M}\sum_{j=1}^{N_i} S\left(\tilde{\VX}^{(i)}_{j}, \tilde{\VY}^{(i)}_{j}\right),
\label{eq:new-metric}
\end{align}
and
\begin{align}
S(\tilde{\VX}^{(i)}_{j}, \tilde{\VY}^{(i)}_{j})= \frac{1}{K_i}\sum_{k=1}^{K_i} \left( \tilde{p}_{k}^{(i)}l\left(\tilde{\Vx}^{(i)}_{j,k},\tilde{\Vy}^{(i)}_{j,k}\right)+ \tilde{q}_{k}^{(i)}s\left(\tilde{\Vx}^{(i)}_{j,k},\tilde{\Vy}^{(i)}_{j,k}\right) \right),
\label{eq:feature_sim}
\end{align}
where $N=\sum_{i=0}^{M}N_i$, $\tilde{q}^{(i)}_k = 1 - \tilde{p}^{(i)}_k$, and $\tilde{p}^{(i)}_k$ is the texture probability of the $k$th patch viewed at the $i$th scale. $l(\cdot)$ and $s(\cdot)$ are defined in Eq. \eqref{eq:lssim}. A-DISTS ranges from zero to one, with a higher value indicating poorer predicted quality.

\section{Experiments}
In this section, we first compare the proposed A-DISTS with a set of full-reference IQA models in term of quality assessment on traditional and novel algorithm-dependent distortions. We then compare A-DISTS against a smaller set of top-performing models in optimization of image super-resolution methods. 

\begin{table*}[t]
  \centering
  \caption{Performance comparison of A-DISTS against twelve IQA models on four standard IQA databases. Larger PLCC, SRCC, and KRCC numbers represent better performance, with a maximum value of one. Top-$2$ results are highlighted in bold.}
  \setlength{\tabcolsep}{2.5mm}{
    \begin{tabular}{lcccccccccccc}
    \toprule
      \multirow{2}{*}[-3pt]{Method} & \multicolumn{3}{c}{LIVE~\cite{LIVE}}&\multicolumn{3}{c}{CSIQ~\cite{larson:011006}}&\multicolumn{3}{c}{TID2013~\cite{Ponomarenko201557}}&\multicolumn{3}{c}{KADID~\cite{lin2019kadid}}\\ 
      \cmidrule(lr){2-4} \cmidrule(lr){5-7} \cmidrule(lr){8-10} \cmidrule(lr){11-13} 
      &PLCC & SRCC & KRCC  & PLCC & SRCC & KRCC  & PLCC & SRCC & KRCC  & PLCC & SRCC & KRCC  \\ \hline 
     PSNR  & 0.865 & 0.873 & 0.680  & 0.819 & 0.810 & 0.601  & 0.677 & 0.687 & 0.496 & 0.675 & 0.676 & 0.488\\
     SSIM~\cite{wang2004image}  & 0.937 &0.948 & 0.796 & 0.852 & 0.865 & 0.680 & 0.777 & 0.727 & 0.545 & 0.717 & 0.724 &  0.537\\
     MS-SSIM~\cite{wang2003multiscale}  &  0.940 & 0.951 & 0.805  &  0.889 & 0.906 & 0.730 & 0.830 & 0.786 & 0.605 & 0.820 & 0.826 & 0.635\\
     VIF~\cite{sheikh2006image}  & 0.960 & 0.964 & 0.828 & 0.913 & 0.911 & 0.743 &  0.771   & 0.677 &  0.518 & 0.687 &    0.679 & 0.507\\
     MAD~\cite{larson:011006}  & \textbf{0.968} & \textbf{0.967} & \textbf{0.842}  & \textbf{0.950} & \textbf{0.947} & \textbf{0.797} & 0.827 & 0.781 & 0.604 & 0.799 & 0.799 & 0.603\\
     FSIM$_\mathrm{c}$~\cite{zhang2011fsim}  & \textbf{0.961} &\textbf{0.965} & \textbf{0.836} & 0.919 & 0.931 & 0.769  & \textbf{0.877} & 0.851 & 0.667 & 0.850 & 0.854 & 0.665\\
     GMSD~\cite{xue2014gradient}  & 0.957 & 0.960 & 0.827 & \textbf{0.945}  &  \textbf{0.950} & \textbf{0.804} & 0.855 & 0.804 & 0.634 & 0.845 & 0.847 & 0.664\\
     VSI~\cite{zhang2014vsi}  & 0.948 & 0.952 & 0.806 & 0.928 & 0.942 & 0.786  & \textbf{0.900} & \textbf{0.897} & \textbf{0.718} & 0.877 & 0.879 & 0.691\\
     NLPD~\cite{laparra2016perceptual}  & 0.932 & 0.937 & 0.778 & 0.923 & 0.932 & 0.769 & 0.839 &  0.800  & 0.625 & 0.811 & 0.812 & 0.623\\
     \hline
     PieAPP~\cite{prashnani2018pieapp}  & 0.908 & 0.919 & 0.750 & 0.877 & 0.892 & 0.715 & 0.859 & \textbf{0.876} & \textbf{0.683} & 0.836 & 0.836 & 0.647\\
     LPIPS~\cite{zhang2018unreasonable}  & 0.934 & 0.932 & 0.765 & 0.896 & 0.876 & 0.689 & 0.749 & 0.670 & 0.497 & 0.839 & 0.843 & 0.653\\
     DISTS~\cite{ding2020dists} & 0.954 & 0.954 & 0.811 & 0.928 & 0.929 & 0.767& 0.855 & 0.830 & 0.639 & \textbf{0.886} &  \textbf{0.887} & \textbf{0.709}\\
     \hline
     A-DISTS (ours) & 0.954 & 0.955 & 0.812 &0.944 & 0.942 & 0.796 & 0.861  & 0.836 & 0.642 &\textbf{0.891} & \textbf{0.890} & \textbf{0.715} \\
    \bottomrule
    \end{tabular}
    }
  \label{tab:4_iqa_database}
\end{table*}

\begin{table*}
\newcommand{\tabincell}[2]{\begin{tabular}{@{}#1@{}}#2\end{tabular}}
\centering
\caption{2AFC score comparison of IQA models on BAPPS and Ding20. It is computed by $r\hat{r}+(1-r)(1-\hat{r})$, where $r$ is the ratio of human votes and $\hat{r}\in\{0,1\}$ is the preference of an IQA model. A higher score indicates better performance.}
\setlength{\tabcolsep}{1.4mm}{
  \begin{tabular}{lcccccccccc}
  \toprule
   \multirow{2}{*}[-6pt]{IQA Model} & \multicolumn{5}{c}{BAPPS~\cite{zhang2018unreasonable}}& \multicolumn{5}{c}{Ding20~\cite{ding2020optim}} \\
   \cmidrule(lr){2-6} \cmidrule(lr){7-11}
   & \tabincell{c}{Color-\\ization}  & \tabincell{c}{Video\\ deblurring} & \tabincell{c}{Frame\\interpolation} & \tabincell{c}{Super-\\resolution} & All &
   Denoising & Deblurring & \tabincell{c}{Super-\\resolution} & Compression & All\\ \hline 
   Human & 0.688 & 0.671 & 0.686 & 0.734 & 0.695 & 0.761 & 0.843 & 0.833 & 0.891 & 0.832\\ \hline 
   PSNR  & 0.624 & 0.590 & 0.543 & 0.642 & 0.614 & 0.627 & 0.518 & 0.612 & 0.689 & 0.612\\
   SSIM~\cite{wang2004image}  & 0.522 & 0.583 & 0.548 & 0.613 & 0.617 & \textbf{0.636} & 0.575 & 0.599 & 0.649 & 0.615\\
   MS-SSIM~\cite{wang2003multiscale} & 0.522 & 0.589 & 0.572 & 0.638 & 0.596 & 0.623 & 0.568 & 0.655 & 0.665 & 0.628\\
   VIF~\cite{sheikh2006image}  & 0.515 & 0.594 & 0.597 & 0.651 & 0.603 & 0.589 & 0.607 & 0.655 & 0.540 &0.598\\
   MAD~\cite{larson:011006}  & 0.490 & 0.593 & 0.581 & 0.655 & 0.599 & 0.624 & 0.671 & 0.681 & 0.651 & 0.657\\
   FSIM$_\mathrm{c}$~\cite{zhang2011fsim}  & 0.573 & 0.590 & 0.581 & 0.660 & 0.615 & 0.522 & 0.490 & 0.525 & 0.563 & 0.525\\
   GMSD~\cite{xue2014gradient}  & 0.517 & 0.594 & 0.575 & 0.676 & 0.613 & 0.417 & 0.454 & 0.469 & 0.567 & 0.477\\
   VSI~\cite{zhang2014vsi}  &  0.597 & 0.591 & 0.568 & 0.668 & 0.622 & 0.518 & 0.470 & 0.487 & 0.576 & 0.513\\
   NLPD~\cite{laparra2016perceptual}  & 0.528 & 0.584 & 0.552 & 0.655 & 0.600 & 0.622 & 0.514 & 0.629 & 0.652 & 0.604\\ \hline
   PieAPP~\cite{prashnani2018pieapp} & 0.594 &  0.582 & 0.598 & 0.685 & 0.626 & 0.625 & 0.734 &  0.744 & 0.822 & 0.732\\
   LPIPS~\cite{zhang2018unreasonable}  & \textbf{0.625} & \textbf{0.605} & \textbf{0.630} & 0.705 & 0.641 & \textbf{0.657} & 0.788 & \textbf{0.768} & \textbf{0.834} & \textbf{0.761} \\
   DISTS~\cite{ding2020dists} & \textbf{0.627} &  0.600 &  \textbf{0.625} & \textbf{0.710} & \textbf{0.651} & 0.602 &  \textbf{0.790} &   0.704 &  0.833 & 0.725\\ \hline
   A-DISTS (ours)  & 0.621 & \textbf{0.602} & 0.616 & \textbf{0.708} & \textbf{0.642} & 0.629 & \textbf{0.792} & \textbf{0.781} & \textbf{0.846} & \textbf{0.763} \\
  \bottomrule
  \end{tabular}
  }
\label{tab:2afc_bapps_our}
\end{table*}

\subsection{Performance on Quality Assessment} 
We use three criteria to evaluate the quality assessment performance, including the Pearson linear correlation coefficient (PLCC), the Spearman rank correlation coefficient (SRCC), and the Kendall rank correlation coefficient (KRCC). A four-parameter function is fitted to compensate for a smooth nonlinear relationship when computing PLCC \cite{ding2020dists}. Following the practice of SSIM and DISTS, A-DISTS also re-scales the smaller dimension of the test images to $256$ pixels. The size of the sliding window in A-DISTS is $21\times 21$ with a stride of one.
We compare A-DISTS with twelve full-reference IQA methods, including nine knowledge-driven models - PSNR, SSIM~\cite{wang2004image}, MS-SSIM~\cite{wang2003multiscale},  VIF~\cite{sheikh2006image}, MAD~\cite{larson:011006}, FSIM$_\mathrm{c}$~\cite{zhang2011fsim}, GMSD~\cite{xue2014gradient}, VSI~\cite{zhang2014vsi}, NLPD~\cite{laparra2016perceptual} and three data-driven CNN-based models - PieAPP~\cite{prashnani2018pieapp}, LPIPS~\cite{zhang2018unreasonable}, DISTS~\cite{ding2020dists}. The implementations of all methods are obtained from the respective authors. 
  
We first show the correlation results on four traditional IQA databases LIVE~\cite{LIVE}, CSIQ~\cite{larson:011006}, TID2013~\cite{Ponomarenko201557}, and KADID~\cite{lin2019kadid}, consisting of traditional distortion types. The former three datasets have been publicly available for many years, and have been extensively re-used throughout the model design process. Released in 2019, KADID is currently the largest human-rated IQA dataset with $81$ original images and $10,125$ distorted images, respectively. From Table \ref{tab:4_iqa_database}, we find that knowledge-driven models (such as MAD \cite{larson:011006} and FSIM$_\mathrm{c}$ \cite{zhang2011fsim}) generally perform better on LIVE, CSIQ, and TID2013, but underperform DISTS and the proposed A-DISTS on KADID. This indicates a potential overfitting issue in the field of IQA, arising from excessive hyperparameter tuning and computational module selection. Ding \etal~\cite{ding2020optim} obtained a similar result in the context of perceptual optimization. Compared to DISTS as the closest alternative, A-DISTS obtains clear perceptual gains on all four databases measured by all three correlation measures.

\begin{table*}
\centering
\caption{Performance comparison of IQA models on four image restoration databases.}
\setlength{\tabcolsep}{2.3mm}{
  \begin{tabular}{lcccccccccccc}
  \toprule
    \multirow{2}{*}[-3pt]{IQA Model} & 
    \multicolumn{3}{c}{Liu13~\cite{liu2013no} (Deblurring)} &
    \multicolumn{3}{c}{Ma17~\cite{ma2017learning} (Super-resolution)}&
    \multicolumn{3}{c}{Min19~\cite{min2019quality} (Dehazing)} &
    \multicolumn{3}{c}{Tian19~\cite{tian2018benchmark} (Rendering)}\\ 
    \cmidrule(lr){2-4}\cmidrule(lr){5-7}
    \cmidrule(lr){8-10}\cmidrule(lr){11-13}
    &PLCC & SRCC & KRCC  & PLCC & SRCC & KRCC  & PLCC & SRCC & KRCC  & PLCC & SRCC & KRCC \\ \hline 
  PSNR  & 0.807 & 0.803 & 0.599 & 0.611 & 0.592 & 0.414 & 0.754 & 0.740 & 0.555 &  0.605 & 0.536 & 0.377\\
  SSIM~\cite{wang2004image}  & 0.763 & 0.777 & 0.574 & 0.654 & 0.624 & 0.440 & 0.715 & 0.692 & 0.513 & 0.420 & 0.230 & 0.156 \\
  MS-SSIM~\cite{wang2003multiscale} & 0.899 & 0.898 & 0.714 & 0.815 & 0.795 & 0.598 & 0.699 & 0.687 & 0.503 & 0.386 & 0.396 & 0.264\\
  VIF~\cite{sheikh2006image}  & 0.879 & 0.864 & 0.672 & 0.849 & 0.831 & 0.638 & 0.740 & 0.667 & 0.504 & 0.429 & 0.259 & 0.173 \\
  MAD~\cite{larson:011006}  & 0.901 & 0.897 & 0.714 & 0.873 & 0.864 & 0.669 & 0.543 & 0.605 & 0.437 & 0.690 & 0.622 & 0.441\\
  FSIM$_\mathrm{c}$~\cite{zhang2011fsim}  & 0.923 & 0.921 & 0.749 & 0.769 & 0.747 & 0.548 & 0.747 &  0.695 & 0.515 & 0.496 & 0.476 & 0.324 \\
  GMSD~\cite{xue2014gradient} & 0.927 & 0.918 & 0.746 & 0.861 & 0.851 & 0.661 & 0.675 &  0.663 & 0.489 & 0.631 & 0.479 & 0.329 \\
  VSI~\cite{zhang2014vsi}  & 0.919 & 0.920 & 0.745 & 0.736 & 0.710 & 0.514 & 0.730 & 0.696 & 0.511 & 0.512 & 0.531 & 0.363 \\
  NLPD~\cite{laparra2016perceptual}  & 0.862 & 0.853 & 0.657 & 0.749 & 0.732 & 0.535 & 0.616 & 0.608 & 0.442 & 0.594 & 0.463 & 0.316\\ \hline
  PieAPP~\cite{prashnani2018pieapp}  & 0.752 & 0.786 & 0.583 & 0.791 & 0.771 & 0.591 & 0.749 & 0.725 & 0.547 & 0.352 & 0.298 & 0.207\\
  LPIPS~\cite{zhang2018unreasonable} & 0.853 & 0.867 & 0.675 & 0.809 & 0.788 & 0.687 & \textbf{0.825} & 0.777 & 0.592 & 0.387 & 0.311 & 0.213 \\
  DISTS~\cite{ding2020dists}  & \textbf{0.940} & \textbf{0.941} & \textbf{0.784} &  \textbf{0.887} & \textbf{0.878} &  \textbf{0.697} & 0.816  & \textbf{0.789} &  \textbf{0.600} & \textbf{0.694} & \textbf{0.671} & \textbf{0.485} \\ \hline
  A-DISTS (ours) & \textbf{0.943} & \textbf{0.944} & \textbf{0.788} & \textbf{0.905} & \textbf{0.892} & \textbf{0.715} & \textbf{0.831} & \textbf{0.801} & \textbf{0.616} & \textbf{0.705}  & \textbf{0.686} & \textbf{0.499} \\
  \bottomrule
  \end{tabular}
  }
\label{tab:srcc_restoration}
\end{table*}

We then compare A-DISTS against the twelve full-reference models on two databases - BAPPS~\cite{zhang2018unreasonable} and Ding20~\cite{ding2020optim}, composed of processed images by real-world image processing systems. BAPPS~\cite{zhang2018unreasonable} is a large-scale patch similarity database with $26,904$ image pairs generated by image colorization, video deblurring, frame interpolation, and super-resolution algorithms. Ding20~\cite{ding2020optim} is a byproduct of a perceptual optimization experiment with $880$ image pairs generated from four low-level vision tasks - image denoising, deblurring, super-resolution, and compression.  Since the human opinions are collected in the two-alternative forced choice (2AFC) experiments, the 2AFC score \cite{zhang2018unreasonable}, which quantifies the consistency of model predictions relative to human opinions, is employed as the evaluation criterion. Results in Table \ref{tab:2afc_bapps_our} show that A-DISTS without reliance on human perceptual scores achieves comparable performance to LPIPS, but is slightly inferior to DISTS on BAPPS. We attribute this to the small patch size (\ie, $64\times64$) of BAPPS, rendering local computation in  A-DISTS less effective. For the images with relatively large size in Ding20, A-DISTS outperforms DISTS and the other models.

We also test A-DISTS on another four publicly available image restoration databases with human judgements:  Liu13~\cite{liu2013no}, Ma17~\cite{ma2017learning}, Min19~\cite{min2019quality}, and Tian19 \cite{tian2018benchmark}, including $1,200$ motion-deblurred images, $1,620$ super-resolved images, $600$ dehazed images, and $140$ rendered images based on depth information, respectively. Table~\ref{tab:srcc_restoration} shows the correlation results, where one can observe that A-DISTS is best at explaining human data in these datasets.

In summary, the proposed A-DISTS achieves better correlation performance than DISTS on all ten databases, except for the patch similarity dataset - BAPPS. This provides strong justifications of the key modifications in A-DISTS: structure and texture separation and locally adaptive weighting.

\subsection{Performance on Perceptual Optimization} \label{sec:sr}
The application scope of objective IQA models is far beyond evaluating image processing algorithms; they can be used as objectives to guide the algorithm design and optimization. In this subsection, we test the gradient-based optimization performance of A-DISTS against four competing models - MAE, MS-SSIM \cite{wang2003multiscale}, LPIPS \cite{zhang2018unreasonable}, and DISTS \cite{ding2020dists} in the context of single image super-resolution. We exclude the rest IQA models in Table \ref{tab:4_iqa_database} because they have been empirically shown less competitive on this task \cite{ding2020optim}.

Single image super-resolution aims to generate a high-resolution (HR) and high-quality image from a low-resolution (LR) one. In recent years, DNN-based methods \cite{edsr17,wang2018esrgan,zhang2019ranksrgan,zhang2020deep} have achieved dominant performance on this task. Here, we adopt the Residual in Residual Dense Block (RRDB) network proposed in \cite{wang2018esrgan} as the backbone to construct our super-resolution algorithms. Training is performed by optimizing a given IQA  model:
\begin{align}
  \ell(\bm{\phi})= D\left(f(
  \VX_l;\bm{\phi}), \VX_h\right),
\label{eq:loss_sr}
\end{align}
where $f(\cdot;\bm{\phi})$ denotes the RRDB network parameterized by a vector $\bm{\phi}$. $\VX_h\in\mathbb{R}^{K}$ is the ground-truth HR image, $\VX_l\in\mathbb{R}^{\left\lfloor\frac{K}{4^2}\right\rfloor}$ is the input LR image down-sampled by a factor of $4$. $D$ represents the IQA metric, with a lower value indicating higher predicted quality.

\begin{figure}
\centering
  \includegraphics[height=0.89\linewidth]{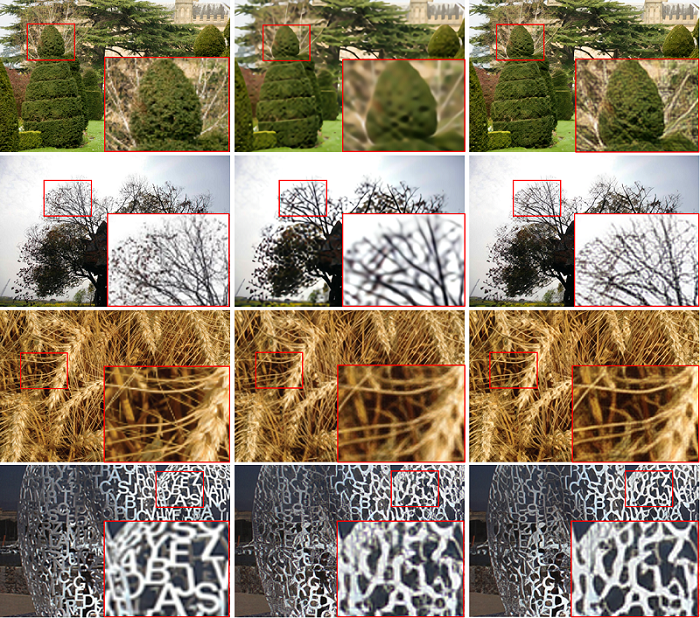}
  \caption{Sample image pairs in our debiased subjective quality assessment. First column: Reference images. Middle column (form top to bottom): Super-resolved images optimized for MAE, MS-SSIM, LPIPS and DISTS, respectively. Last column : Super-resolved images optimized for A-DISTS. See text for more details on image selection.}
\label{fig:pairs_example}
\end{figure}

\begin{figure*}[t]
\centering
  \subfloat{\includegraphics[height=0.1428\linewidth]{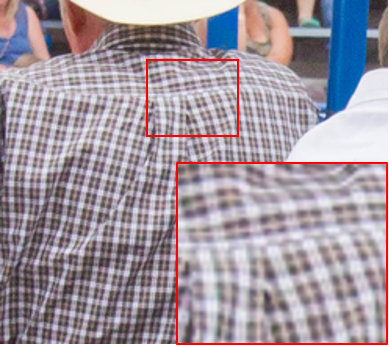}} \hskip.3em
  \subfloat{\includegraphics[height=0.1428\linewidth]{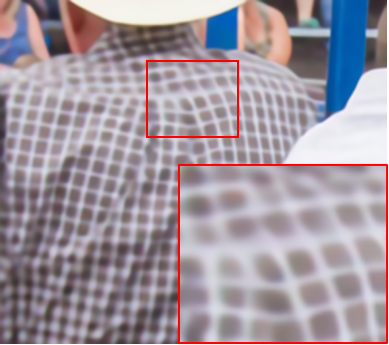}} \hskip.3em
  \subfloat{\includegraphics[height=0.1428\linewidth]{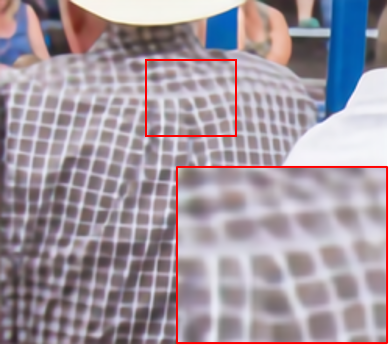}} \hskip.3em
  \subfloat{\includegraphics[height=0.1428\linewidth]{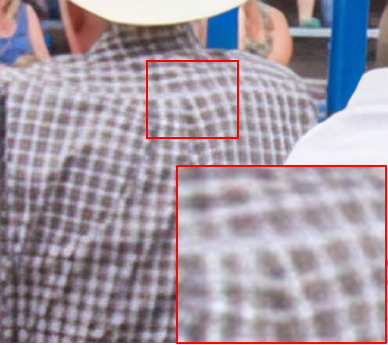}} \hskip.3em
  \subfloat{\includegraphics[height=0.1428\linewidth]{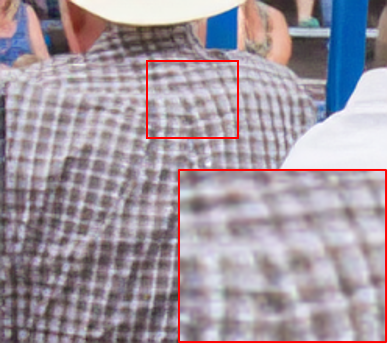}} \hskip.3em
  \subfloat{\includegraphics[height=0.1428\linewidth]{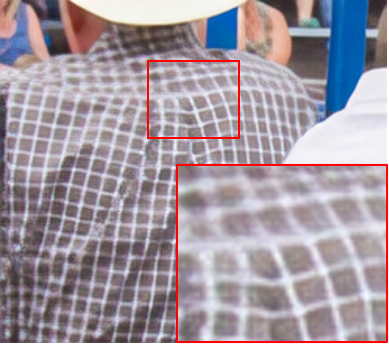}} \\  \vspace{-1em}
  \subfloat{\includegraphics[height=0.1438\linewidth]{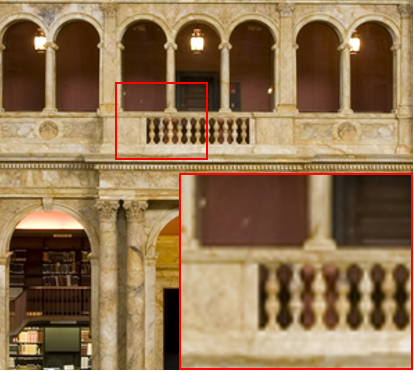}} \hskip.3em
  \subfloat{\includegraphics[height=0.1438\linewidth]{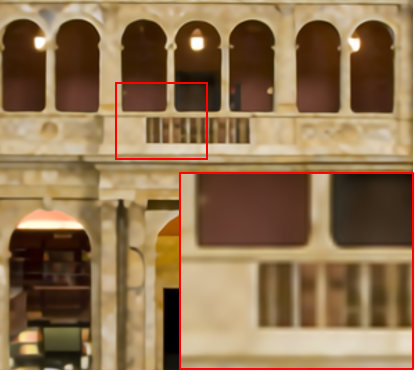}} \hskip.3em
  \subfloat{\includegraphics[height=0.1438\linewidth]{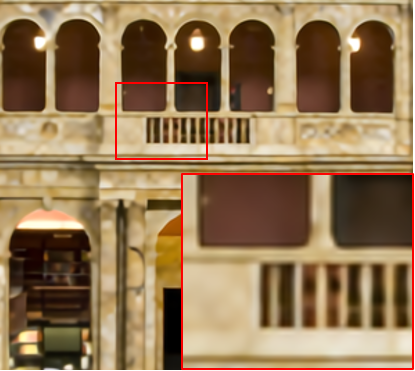}} \hskip.3em
  \subfloat{\includegraphics[height=0.1438\linewidth]{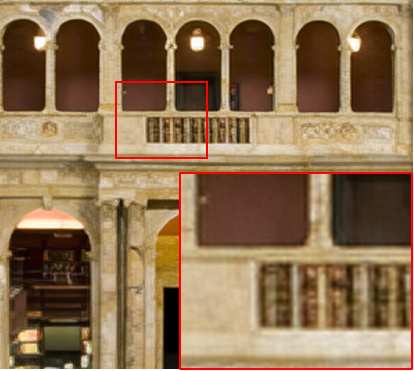}} \hskip.3em
  \subfloat{\includegraphics[height=0.1438\linewidth]{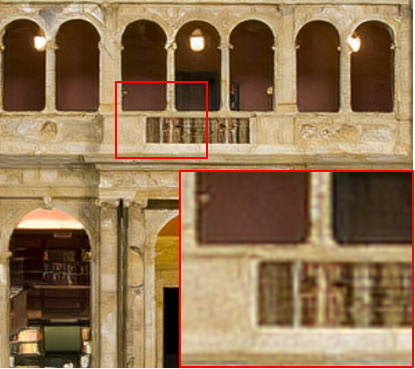}} \hskip.3em
  \subfloat{\includegraphics[height=0.1438\linewidth]{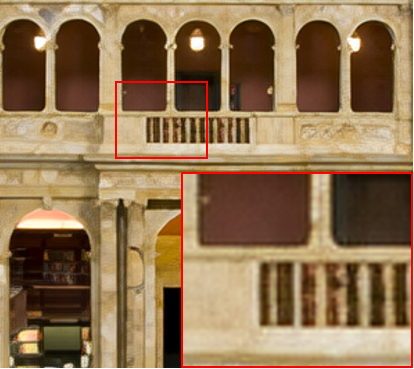}} \\  \vspace{-1em}
  \addtocounter{subfigure}{-12} 
  \subfloat[Original]{\includegraphics[height=0.1436\linewidth]{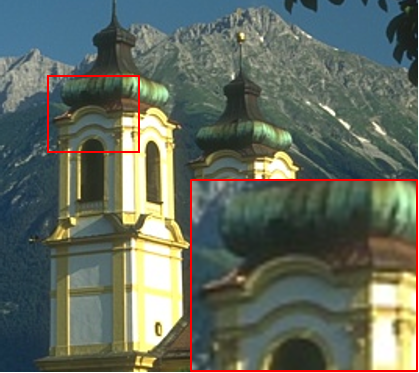}} \hskip.295em
  \subfloat[MAE]{\includegraphics[height=0.1436\linewidth]{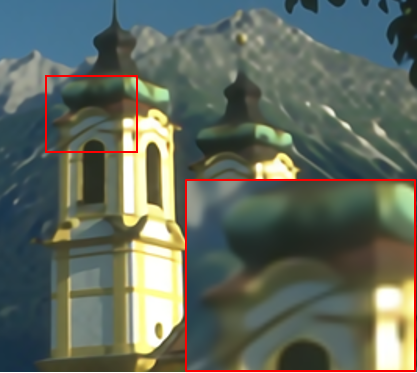}} \hskip.3em
  \subfloat[MS-SSIM]{\includegraphics[height=0.1436\linewidth]{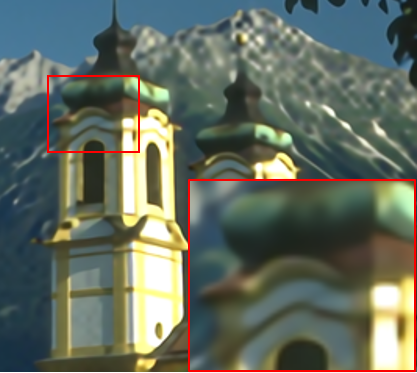}} \hskip.3em
  \subfloat[LPIPS]{\includegraphics[height=0.1436\linewidth]{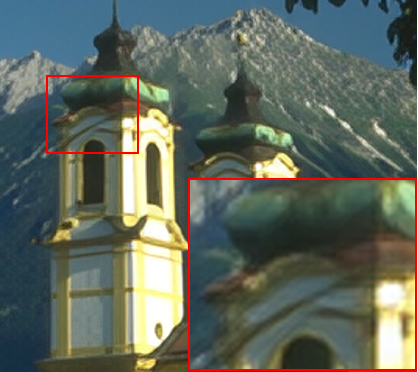}} \hskip.3em
  \subfloat[DISTS]{\includegraphics[height=0.1436\linewidth]{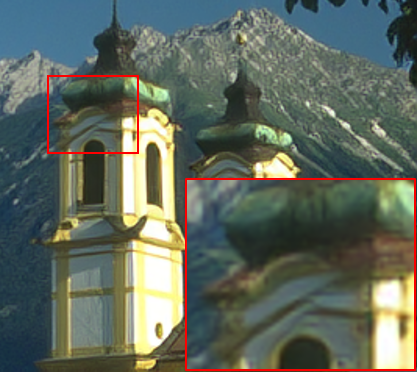}} \hskip.3em
  \subfloat[A-DISTS]{\includegraphics[height=0.1436\linewidth]{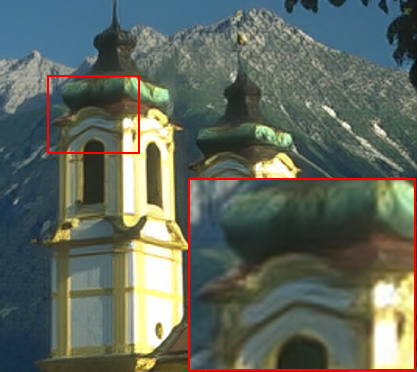}} \\
\caption{Super-resolution results of three example images optimized for different IQA models.}
\label{fig:visual_example}
\end{figure*}

We use the DIV2K database \cite{timofte2017ntire} and the Waterloo Exploration Database \cite{ma2016waterloo} for training and testing, respectively. We generate LR images by downsampling HR images with bicubic interpolation. Following the practice of \cite{ding2020optim}, the model parameters optimized for MAE are employed as the initializations for the networks to be optimized by other models. More training details (\eg, optimizer, learning rate, batch size, etc.) are inherited from \cite{wang2018esrgan}. 
We apply the trained networks to the test images, and conduct a subjective user study for quantitative evaluation. To ensure a fair comparison (\ie, to avoid potential cherry-picking test results), we adopt the debiased subjective assessment method in \cite{cao2021debiased}, which automatically samples a small set of adaptive and diverse test images by solving 
 \begin{align}\label{eq:mad2}
    \VX^{\star} = \argmax_{\VX_l \in \mathcal{X}} ~ \bar{D}(f_i(\VX_l), f_j(\VX_l)), \quad 1 \le i\le j\le 5,
\end{align}
where $\mathcal{X}$ denotes the set of LR images from the Waterloo Exploration Database \cite{ma2016waterloo}. $i$ and $j$ are the algorithm indices. $\bar{D}$ is a measure to approximate the perceptual distance between the super-resolved images $f_i(\VX_l)$ and $f_j(\VX_l)$. We define $\bar{D}$ as the average of two IQA models $D_i$ and $D_j$ used to optimize $f_i$ and $f_j$, respectively\footnote{To compensate for the scale difference, the values of $D_i$ and $D_j$ are mapped to the same MOS scale (\eg, LIVE \cite{LIVE}) by fitting a logistic function.}. By adding a diversity term \cite{cao2021debiased}, we are able to automatically select a small subset of images in $\mathcal{X}$ that best differentiate between two networks $f_{i}$ and $f_{j}$. The comparison is exhausted for all $\binom{5}{2}$ pairs of algorithms. Fig. \ref{fig:pairs_example} shows several sample image pairs in our debiased subjective quality assessment experiment.

\begin{table}
\centering
\caption{Global ranking of the five IQA models for use in optimizing single image super-resolution methods in the debiased subjective testing \cite{cao2021debiased}. A higher ranking score indicates better performance.}
\setlength{\tabcolsep}{1.6mm}{
  \begin{tabular}{lccccc}
  \toprule
  IQA model & MAE & MS-SSIM & LPIPS & DISTS & A-DISTS\\ \hline 
  Ranking score & -1.524 & -1.453 & 0.762 & 0.980 & 1.095 \\
  \bottomrule
  \end{tabular}
  }
\label{tab:ranking}
\end{table}

We employ the 2AFC method for subjective rating.  For each algorithm pair, we set $20$ images according to Eq. \eqref{eq:mad2}. This leads to a total of $\binom{5}{2}\times 20=200$ paired comparisons for $5$ IQA models. Subjects are required to choose the image with higher perceived quality with reference to the ground-truth image. Subjects are allowed to adjust the viewing distance and zoom in/out any part of the images for careful inspection. We gather data from $20$ subjects with general background knowledge of multimedia signal processing. The Bradley–Terry model~\cite{bradley1952rank} is adopted to convert paired comparison results to a global ranking, as shown in Table \ref{tab:ranking}.  We find that the proposed A-DISTS achieves the best perceptual optimization results on average. The ranking of the remaining models is consistent with the conclusions in \cite{ding2020optim}.

Fig. \ref{fig:visual_example} shows three visual examples of super-resolution methods optimized for different IQA models. Like many other studies, we find MAE and MS-SSIM encourage blurry images. The results by DISTS are generally sharper, but appear distortions in structure regions and noise in  texture regions. With locally adaptive structure and texture similarity measurements, A-DISTS generates better visual results with reduced structural artifacts and more plausible textures.


\section{Conclusion and Discussion}%
We have developed a locally adaptive structure and texture similarity index for full-reference IQA. The keys to the success of our approach are 1) the separation of structure and texture across space and scale and 2) the adaptive weighting of quality measurements according to local image content. A-DISTS is free of expensive MOSs for supervised training,  correlates well with human data in standard IQA and image restoration databases, and demonstrates competitive optimization performance for single image super-resolution.

One limitation of the proposed A-DISTS is that the performance on \textit{global} texture-related tasks may be slightly compromised. For example, on the SynTEX database \cite{golestaneh2015effect} for texture similarity, A-DISTS obtains an SRCC of $0.760$ compared to $0.923$ by DISTS. Therefore, a generalized quality measure that translates in a content-dependent way from DISTS to A-DISTS is worth deeper investigation. Nevertheless, as most natural photographic images are made of ``things and stuff'', we believe the proposed A-DISTS holds much promise for use in a wide range of real-world image processing applications. 

\section*{Acknowledgements}
This work was supported in part by Hong Kong RGC Early Career Scheme (No. 21213821 to KDM).

%
\bibliographystyle{ACM-Reference-Format}
\balance

%

\end{document}